\documentclass{revtex4}

\usepackage{amsmath}
\usepackage[dvips]{graphicx}

\def\EQ{\begin{equation}}
\def\EN{\end{equation}}
\def\EQA{\begin{eqnarray}}
\def\ENA{\end{eqnarray}}
\def\ie{{\em i.e. }}

\begin{document}

\title{Intermittency in the homopolar disk-dynamo}
\author{N. Leprovost}
\affiliation{LPS/ENS, 24 rue Lhomond, F-75231 Paris cedex 05}

\author{B. Dubrulle}
\affiliation{DRECAM/SPEC/CEA Saclay, and CNRS (URA2464), F-91190 Gif sur Yvette Cedex, France}

\author{F. Plunian}
\affiliation{LEGI, B.P. 53, 38041 Grenoble Cedex 9, France}

\begin{abstract}
We study a modified Bullard dynamo and show that this system is equivalent to a nonlinear oscillator subject to a multiplicative noise. The stability analysis of this oscillator is performed. Two bifurcations are identified, first towards an \lq\lq intermittent\rq\rq state where the absorbing (non-dynamo) state is no more stable but the most probable value of the amplitude of the oscillator is still zero and secondly towards a \lq\lq turbulent\rq\rq (dynamo) state where it is possible to define unambiguously a (non-zero) most probable value around which the amplitude of the oscillator fluctuates. The bifurcation diagram of this system exhibits three regions which are analytically characterised.
\end{abstract}

\maketitle

\section*{Introduction}
In this paper, we aim to study how a bifurcation is modified when the control parameter fluctuates. This is an important question regarding the problem of self-generation of a magnetic field by means of a conducting fluid, known as the dynamo instability. From the experimental point of view, a working dynamo is one where the fluid is turbulent and thus the velocity (which is the control parameter of the dynamo instability) fluctuates. However, experiments have been designed using the mean-field (mainly time averaged velocity field) and the effect of the fluctuating part of the velocity field still remains an issue \cite{Fauve03,Ponty05}. The main framework to study the effect of a fluctuating velocity field on the dynamo threshold is that of \lq\lq Mean-Field theory\rq\rq (MFT) of dynamo \cite{Krause80,Moffatt78} where the induction equation is supplemented by two terms, an \lq\lq $\alpha$  effect\rq\rq which corresponds to the creation of large-scale magnetic field by helical small-scale motions and a \lq\lq $\beta$  effect\rq\rq which accounts for a turbulent diffusivity. The validity of the mean-field theory has however been questioned due to the statistical nature of the turbulent dynamo theory: due to fluctuations, there may be some additional effects which MFT cannot account for \cite{Hoyng87,Hoyng87b}. Another limitation in the MFT approach  is due to the fact that it is purely linear, disregarding the effect of the magnetic field on the velocity field, and consequently, may overlook some effect due to the interplay between noise and non-linearity, such as stabilisation by noise \cite{Schenzle79,Lucke85}, noise-induced transitions\cite{Horsthemke,VdB94}, stochastic resonance \cite{Gammaitoni98}, etc. 
 
Based on the fact that the fluctuating parameter (the velocity) multiplies the magnetic field, it has been argued that the magnetic field above threshold should be intermittent, exhibiting sporadic growth of the magnetic field \cite{Peffley00,Sweet01}. This feature has been evidenced in the case of a stochastic model of the turbulent dynamo \cite{Leprovost05b} but only under simplified assumptions: zero magnetic viscosity and $\delta$-correlated noise. Consequently, this is not clear if a \lq\lq real dynamo\rq\rq would exhibit intermittency. To shed some light on this problem, we study the much simpler problem of a solid dynamo with a scalar control parameter and we show that this problem can be mapped into that of a nonlinear oscillator with a fluctuating frequency,  a problem that has been widely investigated in the literature. The question of its stability has been questioned both in the linear case \cite{Bourret73} and the nonlinear one \cite{Graham82b,Wiesenfeld82,Lucke85,Mallick04}. We then use known results to draw the stability diagram of this system and show that, in the non-linear case, the system bifurcates towards an intermittent state (characterised by a mostly null field, except for some small amount of time where the field exhibits bursts of activity). We also show that if the intensity of the fluctuations is increased, that particular behaviour can disappear and the field oscillates around a well defined mean value. 

After introducing the disk-dynamo model, we derive the analogy with a noisy oscillator (section \ref{Oscillator}) and in section \ref{Bruitblanc}, both the frontier between the trivial state and the intermittent state (already calculated by \cite{Mallick04}) and between the intermittent state and the oscillatory state are analytically calculated. We also address the effect of noise correlation by means of numerical simulations (section \ref{Correlations}).   

\section{Model}
\label{Model}
\begin{figure}[h]
\begin{center}
\includegraphics[scale=0.8,clip]{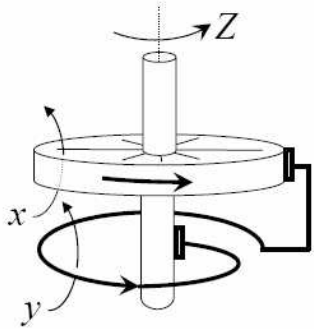}
\caption{\label{Bullard} The homopolar (or Bullard) disk-dynamo.}
\end{center}
\end{figure}

The Bullard (or homopolar) dynamo \cite{Bullard55} is the first example of a magnetic instability triggered by a conductor in motion. The experimental device is depicted on figure \ref{Bullard}: a conducting disk rotates around its axis at the angular velocity $\omega$ and a small magnetic field is applied in the vertical direction inducing a current in the disk from the axis towards the edge of the conducting disk. Then this current flows in a conducting loop which, if orientated in an appropriate way, induced a magnetic field orientated in the same direction as the initial one. Thus an infinitesimal magnetic field can be amplified by this mechanism, leading to a dynamo.
Here we study a modified version of the original Bullard dynamo where an azimuthal current at the edge of the disk and mechanical friction are permitted. The azimuthal current has been introduced by Moffatt \cite{Moffatt79} who showed that the hypothesis of a purely radial current would violate the conservation of magnetic flux in the case of a perfectly conducting disk. According to Hide \cite{Hide95}, neglecting the mechanical friction is unwarranted for it would render the disk-dynamo structurally unstable. Three equations are enough to describe this system \cite{Knobloch81,Plunian98}:
\EQ
\begin{cases}
\label{BullardModif}
\dot{x} = q (y-x)\\
\dot{y} = x Z + mx - (m+1) y\\
\dot{Z} = g [1- (m+1) xy + m x^2] - f Z \\
\end{cases}
\EN
where $Z$ is the dimensionless angular velocity of the disk and $x$ and $y$ are the two components of the magnetic flux, across the disk and across the loop (see \cite{Plunian98} for the precise meaning of the different constants). The two first equations represent the effect of the rotating disk on the magnetic field and the third equation is the evolution of the angular velocity, subject to Joule heating and externally applied torque (the term proportional to $g$) and a mechanical friction (the term proportional to $f$). A linear instability analysis around the solution $Z=Z_0$ (given angular velocity) and $x=y=0$ (zero magnetic flux), proves that the system becomes unstable as soon as $Z_0 > 1$.

\section{A non linear noisy oscillator}
\label{Oscillator}
To mimic the turbulent dynamo problem evoked in the introduction, we make a kind of kinematic approximation: we prescribe a fluctuating velocity field and study the generation of the magnetic field induced by this given velocity. Practically, this accounts to disregard the third equation and study the two first equations only. However we intend to compare the result of this procedure to the numerical analysis of the full system, and while doing such, it could be important to keep track of the non linear term in the third equation (the back reaction of $x$ and $y$ on the velocity $Z$).

Specifically, we decompose the velocity field as a sum of a mean part and a fluctuating part: $Z=Z_0 + \Gamma(t)$. For simplicity, we assume that the fluctuating part of the velocity field is white in time~: $\langle \Gamma(t) \Gamma(t') \rangle = 2D \delta(t-t')$. Applying this procedure to the two first equation of (\ref{BullardModif}), we are left with a linear stochastic system. It is well known that it is difficult to define unambiguously a threshold for this system and it is necessary to account for the non-linearity induced by the third equation. In consequence, we introduce a non-linear term and write $Z=Z_0 + \Gamma(t)+[-(m+1)xy+mx^2] g/f$  to mimic the feedback of the variables $x$ and $y$ on the intensity of the dimensionless angular velocity. 

The constitutive equation of this system can be transformed into the equation describing a nonlinear oscillator in presence of noise and damping:
\EQ
\label{Oscillateur}
\quad \ddot{x} + (1+\beta\lambda x^2) \dot{x} - \alpha x = \xi(t) x - \lambda x^3 \, , \quad \text{with} \quad
\langle \xi(t) \xi(t') \rangle = \Delta \delta(t-t') \, ,
\EN
where the time and the variable $x$ have been adimensionalised by $(m+1+q)^{-1}$. The parameters are:
\EQA
\alpha &=& \frac{q(Z_0-1)}{(m+1+q)^2} \, , \qquad \Delta = \frac{2D q}{(m+1+q)^3} \, , \\ \nonumber
\lambda &=& \frac{q g/f}{(m+1+q)^4} \qquad \text{and} \qquad \beta=\frac{(m+1+q)(m+1)}{q} \, .
\ENA
The equation (\ref{Oscillateur}) is similar to the equation of a Duffing oscillator except for the presence of non-linear terms. The undamped version of this oscillator has first been studied in the context of Anderson localisation problem \cite{Tessieri00} and more recently introducing a damping and studying the linear stability of this oscillator \cite{Mallick04}. Here we focus on the non-linear version of the oscillator and show that the transition observed in the non-linear system is of a very particular nature and that it is possible to identify a series of two bifurcations in this system. To study (\ref{Oscillateur}), we make the following change of variables from the Cartesian coordinates $(x,\dot{x})$ to the polar ones $(r,\theta)$:
\EQ
x = r \cos(\theta) \quad \text{and} \quad \dot{x} = r \sin(\theta) \; .
\EN
Then, the evolution equations for the new variables $r$ and $\theta$ can be derived:
\EQA
\dot{r} &=& r\sin\theta(\cos\theta-\sin\theta)+\alpha r \sin\theta\cos\theta + r \cos\theta\sin\theta \xi(t)  \\ \nonumber
&& \qquad - \lambda r^3 (\cos^3\theta\sin\theta + \beta\cos^2\theta\sin^2\theta) \\ \nonumber
\dot{\theta} &=& -\sin\theta (\cos\theta+\sin\theta) + \alpha\cos^2\theta+ \cos^2\theta \xi(t)
 \\ \nonumber
&&  \qquad -\lambda r^2 (\cos^4\theta + \beta\cos^3\theta\sin\theta)  \; .
\ENA
We are left with a system of two stochastic equations for $r$ and $\theta$. From these equations, using standard techniques of stochastic process \cite{Vkampen81}, it is easy to write a Fokker-Planck equation which governs the evolution of the density probability $H(t,r,\theta)$ to observe a couple $(r,\theta)$ at time t:
\EQ
\label{FPrt}
\partial_t H = \mathcal{L}_r H + \mathcal{L_\theta} H + \mathcal{R} H \; ,
\EN
where $\mathcal{L}_r$ and $\mathcal{L_\theta}$ are two differential operators involving respectively derivative with respect to the radial variable and the angular variable:
\EQA
\mathcal{L}_r H &=& - \bigl[\alpha \sin\theta + \sin\theta(\cos\theta-\sin\theta)+\frac{\Delta}{2} \cos^2\theta(\cos^2\theta-\sin^2\theta) \Bigr] \partial_r(rH) \\ \nonumber
&+& \lambda \bigl(\cos^3\theta\sin\theta+\beta\cos^2\theta\sin^2\theta\bigr) \partial_r(r^3H) + \frac{\Delta}{2} \cos^2\theta\sin^2\theta \partial_r[r\partial_r(rH)] \; , \\ \nonumber
\mathcal{L_\theta} H &=& \partial_\theta\Bigl[\bigl(\sin\theta(\cos\theta+\sin\theta)-\alpha\cos^2\theta+\lambda r^2 (\cos^4\theta+\beta\cos^3\theta\sin\theta)\bigr)H\Bigr] \\ \nonumber
&+&\frac{\Delta}{2}\partial_\theta\bigl[\cos^2\theta\partial_\theta(\cos^2\theta H)\bigr] \; , \\ \nonumber
\mathcal{R} H &=& \Delta \partial_r\partial_\theta\bigl[r\sin\theta\cos^3\theta H\bigr] \; .
\ENA

\subsection{Radial density of probability}
The partial differential equation (\ref{FPrt}) admits simple analytical solutions with separation of radial and angular variables: $H(r,\theta)=P(r)G(\theta)$. Then, if we integrate equation (\ref{FPrt}) relatively to the angular variable, we obtain the following equation for the probability density of the radial variable:
\EQ
\label{FPr}
\partial_t P = a \partial_r \bigl[ r \partial_r ( r P) \bigr] - b \partial_r\bigl[r P] + c\partial_r[r^3 P] \; ,
\EN
with the following parameters expressed through an average over the angular variable ($\langle \bullet \rangle_\theta = \int \bullet \; G(\theta) d\theta$):
\EQA
a &=& \frac{\Delta}{2} \langle \cos^2\theta \sin^2\theta \rangle_\theta \; , \\ \nonumber
b &=& \langle \alpha\sin\theta\cos\theta + \sin\theta(\cos\theta-\sin\theta) + \frac{\Delta}{2} \cos^2\theta (\cos^2\theta-\sin^2\theta)\rangle_\theta \; , \\ \nonumber
c &=& \lambda \langle \cos^3\theta \sin\theta + \beta \cos^2\theta \sin^2\theta \rangle_\theta \; .
\ENA
It is then easy to find a stationary solution of equation (\ref{FPr}):
\EQ
\label{Prstat}
\quad P_s(r) = \frac{1}{Z} r^{b/a-1}\exp [-\frac{c r^2}{2a}] \qquad \text{with} \quad
Z = \frac{1}{2} \Bigl[\frac{2 a}{c}\Bigr]^{b/(2 a)} \Gamma(\frac{b}{2 a}) \; .
\EN
When $b$ is negative, one can check that the distribution (\ref{Prstat}) is not integrable in zero. In that case, the only admissible solution is a Dirac function centred around zero: $P_s(r) = \delta(r)$ which corresponds to a solution $x = y = 0$ at long time. Indeed, one can check that it is always a solution of equation (\ref{Oscillateur}). Then, we can identify two bifurcations whether the control parameter is taken to be the mean value of $r$ or its most probable value. This scenario as already been evidenced in the case of a stochastic modelling of the dynamo effect \cite{Leprovost05b}. To compute the threshold value corresponding to this two bifurcations, one needs to characterise completely the probability density (\ref{Prstat}) and thus to calculate the coefficient defined above. To achieve that, one needs to compute the stationary probability density of the angular variable.

\subsection{Angular distribution of probability}
To express the coefficient $a$, $b$ and $c$, we have to compute the distribution probability of the variable $\theta$. This can be done by averaging equation (\ref{FPrt}) over the $r$ variable. This leads to the following equation:
\EQA
\label{FPt}
\partial_t G &=& - \partial_\theta\bigl[(\alpha \cos^2\theta - \sin\theta(\cos\theta+\sin\theta))G\bigr] + \frac{\Delta}{2} \partial_\theta \bigl[\cos^2\theta \partial_\theta (\cos^2\theta G)\bigr] \\ \nonumber
&+& m \partial_\theta \bigl[(\cos^4\theta+\beta\cos^3\theta\sin\theta)G\bigr] \; ,
\ENA
where we set $m = \lambda \int r^2 P_s(r) dr = b \lambda / c$. It is interesting to notice that this coefficient does not depend on $\lambda$ and thus that the probability distribution does not depend on the intensity of the non-linear terms. It is convenient to solve the preceding equation to introduce the variable $z = tan\theta = \dot{x}/x$ which leads to the following Fokker-Planck equation for the probability distribution $I(z,t)$:
\EQ
\partial_t I = - \partial_z \Bigl[\Bigl(\frac{\alpha}{1+z^2} - \frac{z}{1+z^2} + m \frac{1+\beta z}{(1+z^2)}\Bigr) I \Bigr] + \frac{\Delta}{2} \partial_z^2 I \; .
\EN
This equation can be solved in the stationary case ($\partial_t I = 0$) to obtain:
\EQA
\label{SolutionI}
I(z) &=& \frac{1}{Z} \int_{-\infty}^z \exp\Bigl[\frac{2}{\Delta}\bigl(\Psi(z) - \Psi(t)\bigr) \Bigr] \, dz \\ \nonumber
\text{with} \quad \Psi(t) &=& \alpha t - \frac{t^2}{2} - \frac{t^3}{3} - m \Bigl[\arctan(t) + \frac{\beta}{2} \log(1+t^2) \Bigr] \; .
\ENA
This expression is identical to that of \cite{Mallick04} except for the last term which comes from the nonlinearity and which is important for the structure of the oscillator bifurcation diagram, as we now discuss.

\section{Bifurcation scenario}
\label{Bruitblanc}
The bifurcation of the noisy oscillator is of a rather complex nature. Indeed, depending on the control parameter, two thresholds can be identified. This is to be contrasted with the case without noise where $\alpha< 0$ corresponds to an absorbing state where the oscillator relax towards $x = 0$ and $\alpha > 0$ where the solution at long time is $x=\sqrt{\alpha/\lambda}$. We now characterise more deeply the nature of the bifurcation when the noise intensity is not zero.

\subsection{$b > 0$: bifurcation towards an intermittent state}
\begin{center}
\begin{figure}[h]
\includegraphics[scale=0.4,clip]{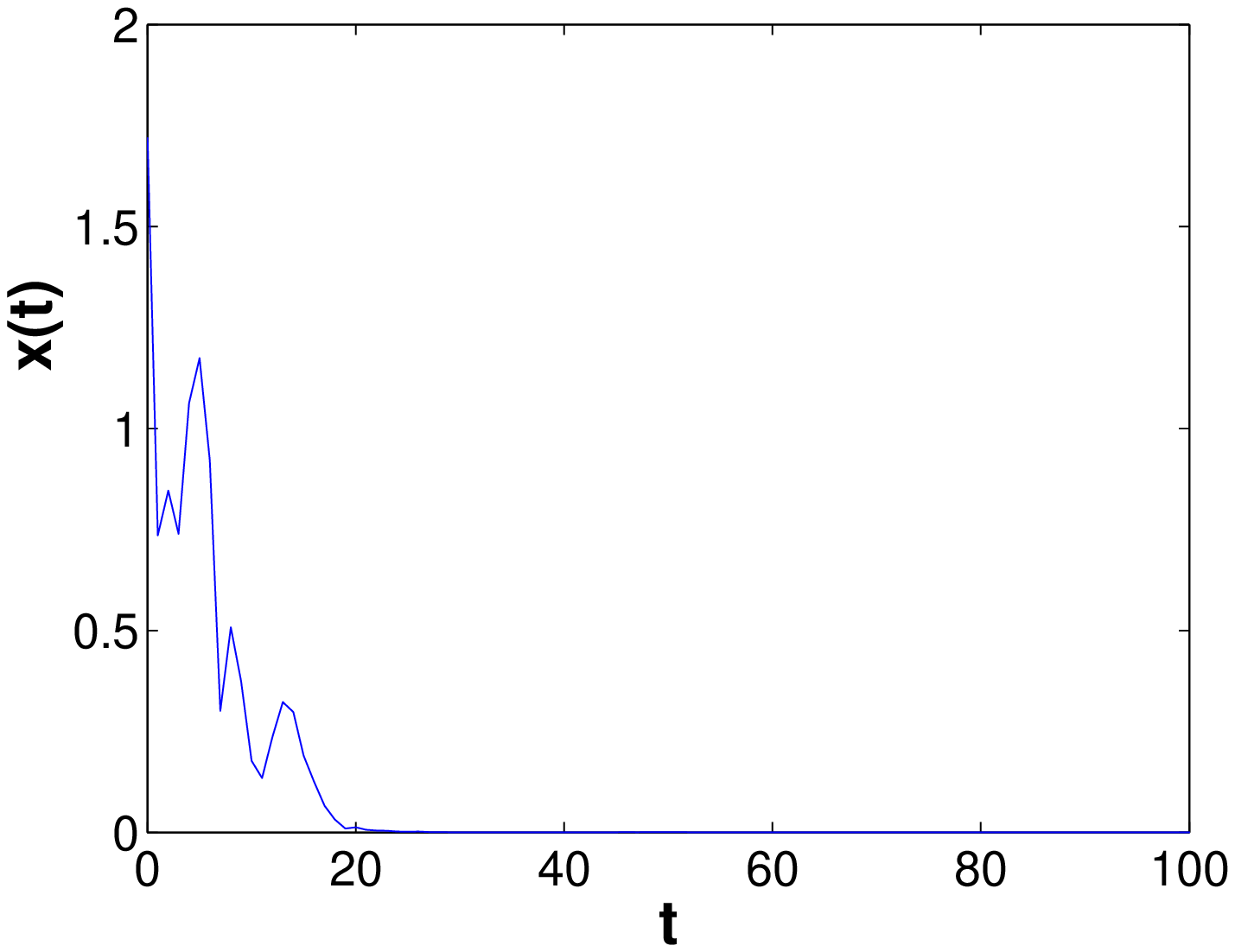}
\includegraphics[scale=0.31,clip]{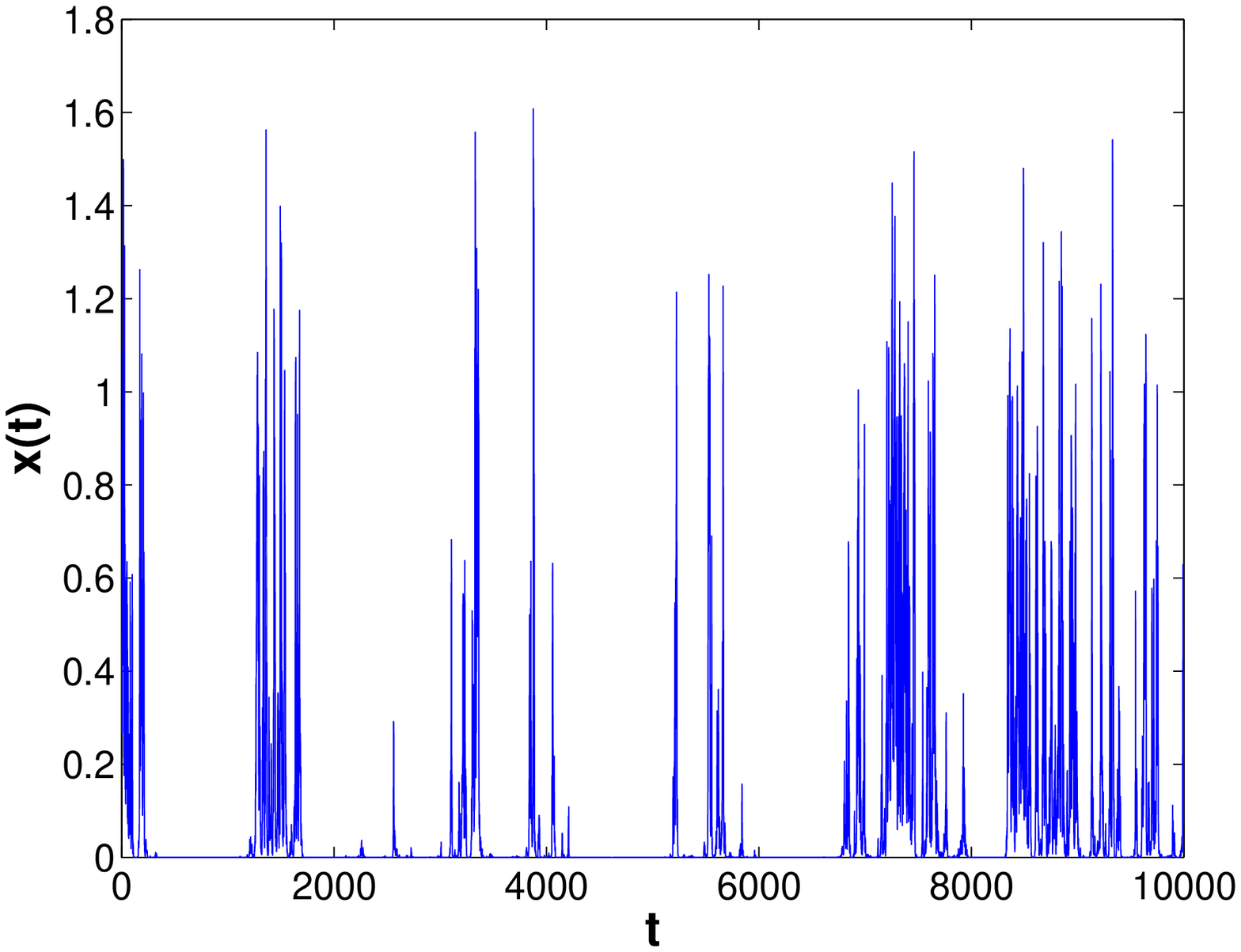}
\caption{\label{TransInt} Result of a numerical simulation of system (\ref{Oscillateur}) with $\Delta=2$, $\lambda=1$, $\beta=3$. The left panel is for $\alpha=-0.2$, corresponding to an absorbing state and the right panel for $\alpha=0.2$, corresponding to an intermittent state.}
\end{figure}
\end{center}

When $b$ becomes positive, the distribution (\ref{Prstat}) is integrable near the origin and can thus represent a meaningful probability distribution function. However, the distribution diverges in zero, an indication that the most probable value is still zero. This is characteristic of an intermittent state where the signal exhibit bursts of activity separated by quiescent epochs where the norm is close to zero. To illustrate such a behaviour, we performed numerical simulations of the equation (\ref{Oscillateur}) and a typical snapshot is shown on figure \ref{TransInt}.
On the left hand side, we see that the variable $x$ after a short transient time (compare the time interval of the two snapshots) relax towards the absorbing state. Increasing the parameter $\alpha$ (keeping the other parameter fixed at the value of figure \ref{TransInt}), one notice that a first bifurcation occurs (around $\alpha=0.17$) which leads to a state where the variable $x$ is most of the time close to zero but exhibits bursts (cf the right hand side of the figure). We call this type of behaviour intermittency.

\begin{center}
\begin{figure}[h]
\includegraphics[scale=0.45,clip]{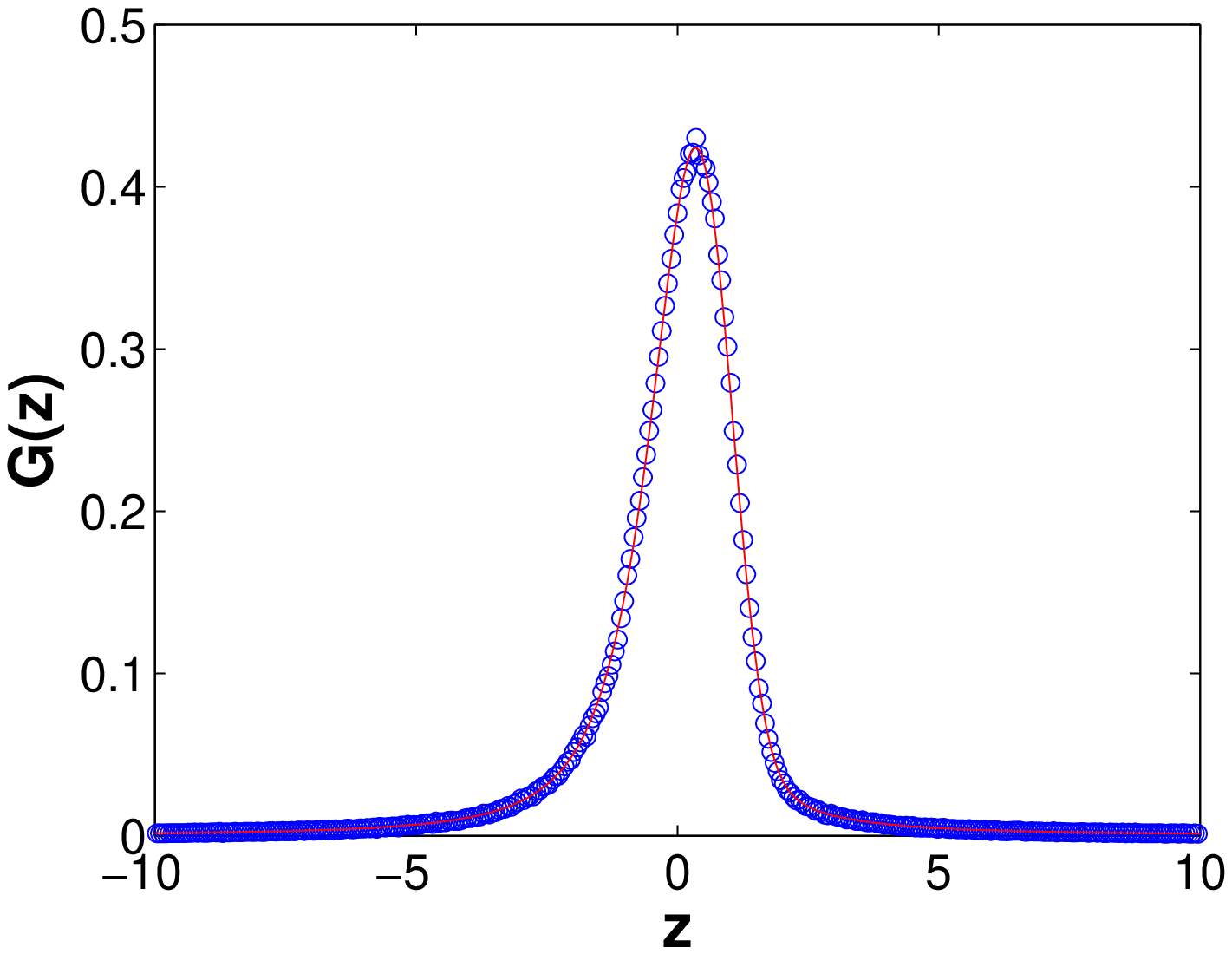}
\includegraphics[scale=0.45,clip]{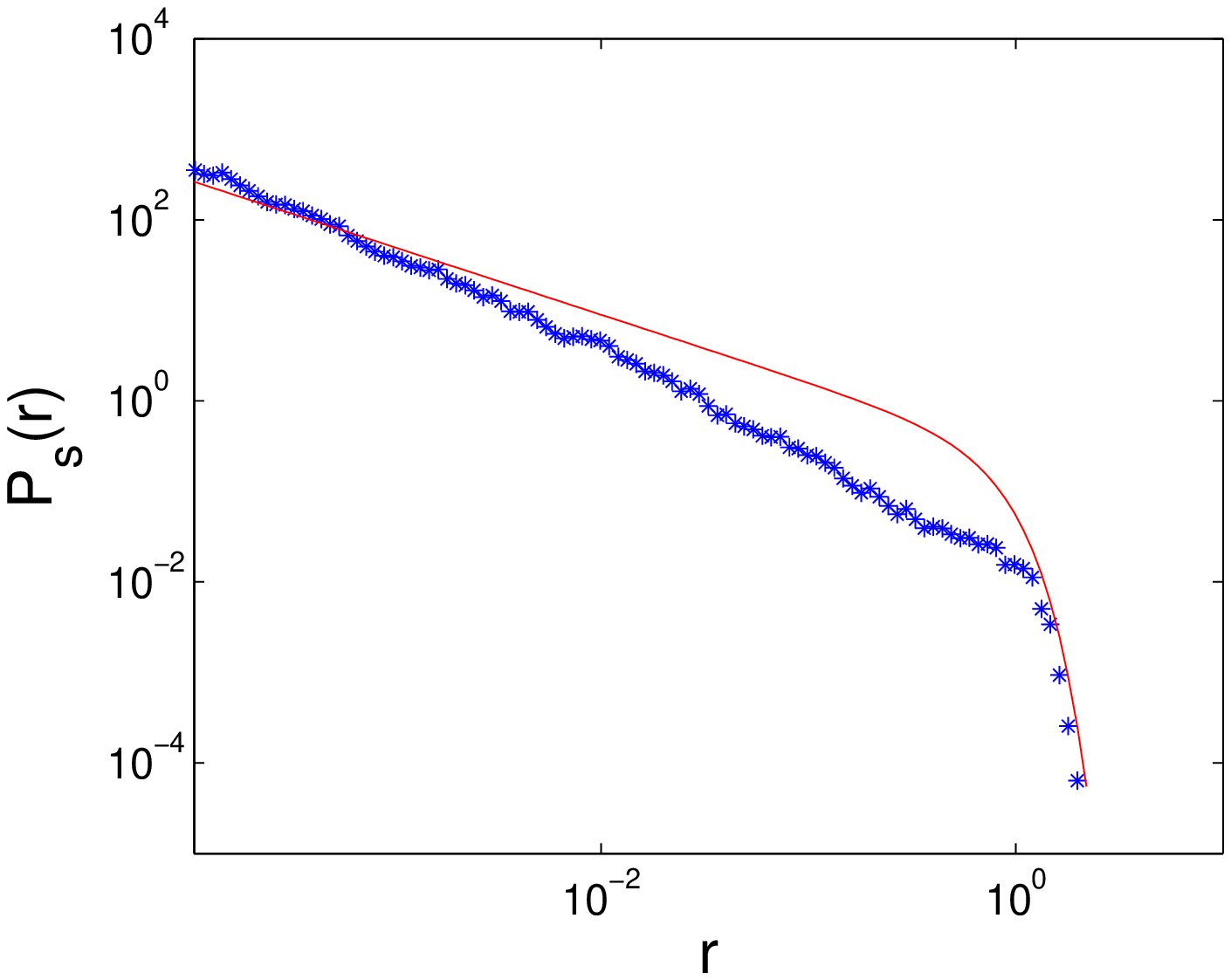}
\caption{\label{PDFInt} Comparison of the simulation with $\Delta=2$, $\lambda=1$, $\beta=3$ and $\alpha=0.2$ corresponding to an intermittent sate and the analytical prediction (in straight line) for the probability density of the variable  $z$ (left hand side) and  $r$ (right hand side in log-log scale). The parameters have been found to be numerically $a$ = 0.14, $b=0.019$, $c=0.47$ and $m=0.04$.}
\end{figure}
\end{center}

To characterise the intermittent state, we use the time series of $x$ to compute the different parameters that appeared in the previous section $a$, $b$, $c$ and $m$. Once this parameters are known, we can compute the probability distribution of the variable $r$ and $z$ given by the expression (\ref{Prstat}) and (\ref{SolutionI}). The theoretical distribution for the angular variable shows a good agreement with the numerical ones as is shown on figure \ref{PDFInt}. Concerning the radial distribution, the agreement is good for low values of $r$ but not for high values. This is because the decoupling approximation is supposed to be relevant for low energies (or low $r$). However, because we are interested in the stability of the $r=0$ fixed point, the approximation is good in the region of interest.

\subsection{$b >a$: bifurcation towards a turbulent state}
\begin{center}
\begin{figure}[h]
\includegraphics[scale=0.45,clip]{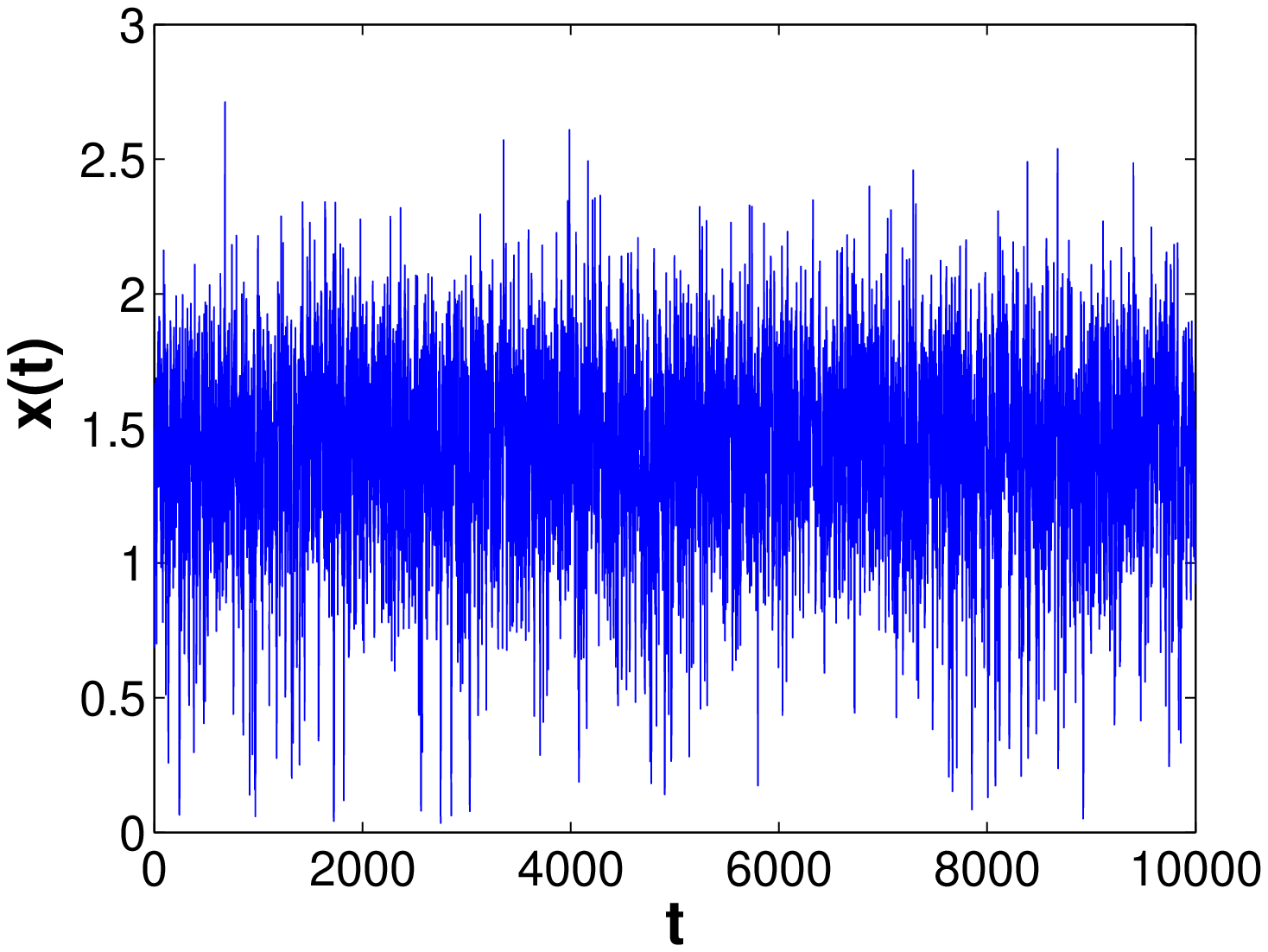}
\includegraphics[scale=0.45,clip]{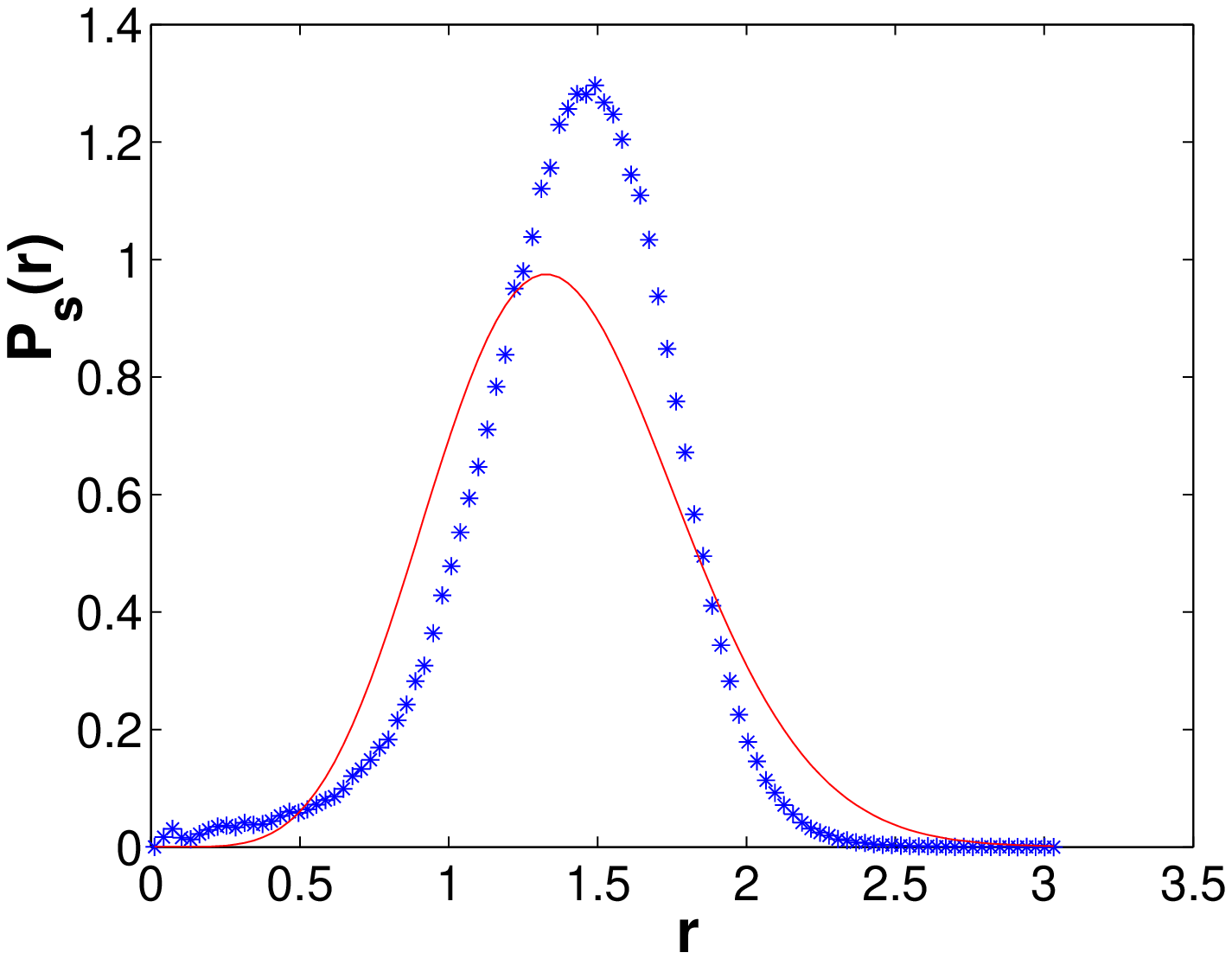}
\caption{\label{PDFturb} Time series (left) of the variable $x$ and probability distribution (right) of the variable $r$ for values of the parameters corresponding to a turbulent state: $\Delta=2$, $\lambda=1$, $\beta=3$ and $\alpha=2$. The numerically calculated value are: $a=0.088$, $b=0.54$ and $c=0.25$.}
\end{figure}
\end{center}

When $b$ becomes larger than $a$, one can easily check that the distribution has a well defined maximum for $r =\sqrt{(b-a)/c}$. A typical example of this behaviour is shown on the left of figure \ref{PDFturb}.
In this new regime, the variable $x(t)$ fluctuates around a well defined most probable value which can be seen both on the time series (left) and on the probability distribution which has a well defined maximum. One sees that the theoretical distribution does not fit the numerical one accurately anymore, a possible indication of the breaking down of the separation hypothesis. This prevents us to make any prediction on the mean value of $x$ in the turbulent regime. However, due to the fact that the theoretical distribution fits well the numerical one in the intermittent regime, it is still possible to calculate the boundary between the intermittent and the turbulent regime and to draw the phase diagram of the system. We will now proceed to this determination.

\subsection{Stability Diagram}
In previous sections, we identified a scenario of two bifurcations for the non linear oscillator: starting from the absorbing state $x=0$, a first bifurcation leads to an {\em intermittent} regime (see figure \ref{PDFInt}) followed by a second bifurcation leading to a {\em turbulent} regime. These two regimes can be identified via the shape of the probability density of the variable $r$, or, equivalently, by means of conditions on the coefficient $a$ and $b$. We now write these conditions in terms of the initial variable $\alpha$, $\Delta$, $\lambda$ and $\beta$, so as to fully characterise the phase space.

For the first bifurcation (towards the intermittent state), we notice that the position of the bifurcation line does not depend on the non-linear terms. Indeed, under this threshold, the only solution is $x =0$ and thus the non linear term is negligible. This has already been stressed by \cite{Mallick04} and we here just outline the main lines of the computation. First, multiplying equation (\ref{FPt}) by $log(1+z^2)$ and averaging over $z$, one finds the following relation:
\EQ
b = \langle z + m \frac{z+\beta z^2}{(1+z^2)^2} \rangle \; ,
\EN
which reduces to $b = \langle z \rangle$ upon discarding the non linear terms. This is the expression for the Lyapunov exponent found by \cite{Tessieri00} in the case of Anderson's localisation problem and later by \cite{Mallick04} when studying the linear version of our oscillator. Using, the expression for the probability distribution of $z$ (\ref{SolutionI}), one may then express the condition of instability $b > 0$ and find the bifurcation line $\alpha^*(\Delta)$.

\begin{center}
\begin{figure}[h]
\includegraphics[scale=0.8,clip]{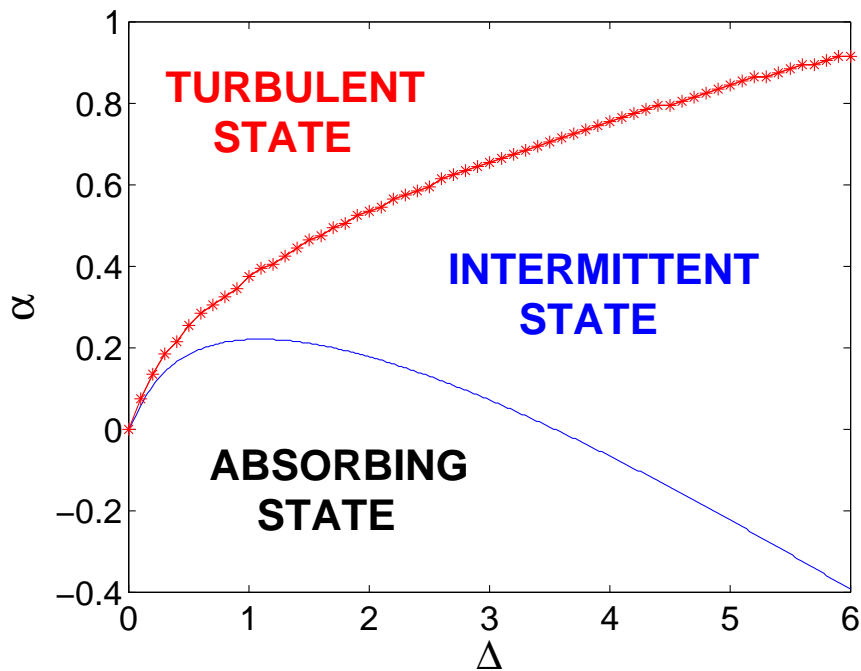}
\caption{\label{DiagrInt} Bifurcation diagram of the noisy oscillator (\ref{Oscillateur}) for $\beta=3$. The full line curve correspond to the condition (\ref{ConditionStab}) and the one with \lq\lq $\star$\rq\rq \, sign to the condition (\ref{ConditionInt}).}
\end{figure}
\end{center}

The condition for the appearance of intermittency can thus be written as $b = \langle z \rangle > 0$. As shown by \cite{Mallick04}, this can be written:
\EQ
\label{ConditionStab}
\int_0^{+\infty} \, du \, \sqrt{u} \exp\Bigl[\frac{2}{\Delta}\Bigl\{(\alpha+\frac{1}{4})u-\frac{u3}{12}\Bigr\} \Bigr] > \int_0^{+\infty} \frac{du}{\sqrt{u}} \exp\Bigl[\frac{2}{\Delta}\Bigl((\alpha+\frac{1}{4})u-\frac{u3}{12}\Bigr) \Bigr] \; .
\EN
The numerical integration of this condition give the bifurcation line $\alpha(\Delta)$ which is drawn on figure \ref{DiagrInt} in full line. This curve delineates the parameter space between the absorbing state (or the \lq\lq no dynamo\rq\rq state), corresponding to $x$ going to zero for long time and the (intermittent and turbulent) dynamo states.

To draw the frontier between these two states, one need to re-express the condition $b > a$ in terms of the initial parameters. This can be written:
\EQ
\label{ConditionInt}
\Bigl\langle z + m \frac{z+\beta z^2}{(1+z^2)^2} \Bigr\rangle > \frac{\Delta}{2} \Bigl\langle \frac{z^2}{(1+z^2)^2} \Bigr\rangle \quad \text{where} \quad m=\frac{b \lambda}{c} = \frac{\Bigl\langle z + m \frac{z+\beta z^2}{(1+z^2)^2} \Bigr\rangle}{\Bigl\langle \frac{z+\beta z^2}{(1+z^2)^2} \bigr\rangle} \; ,
\EN
where the brackets correspond to a mean using the distribution (\ref{SolutionI}). Notice that this condition and the distribution only depend on $\alpha$, $\Delta$, $\beta$ and $m$. Furthermore, it is easy to see that the factor $\lambda$ cancels out in the expression for $m$. As a consequence, the bifurcation line $\alpha(\Delta,\beta)$ between the intermittent and the turbulent state does not depend on the particular value of $\lambda$, \ie on the absolute intensity of the non linear terms. This is a rather remarkable feature of the bifurcation. The result of the numerical integration of the two equations (\ref{ConditionInt}) is shown on the figure \ref{DiagrInt}. One sees that the evolution of this threshold is monotonous: when the noise is increased, the transition from the intermittent to the turbulent state is delayed. On the contrary, the transition from the absorbing state to the intermittent one is first increased for weak intensities of the noise (stabilisation by noise) and then is lowered below its deterministic value for more powerful noise (a reentrant transition that has been pointed out by \cite{Mallick04}).

\section{Influence of the noise correlation}
\label{Correlations}
The next step will be to compare the stability diagram of figure \ref{DiagrInt} with a numerical simulation of the system (\ref{BullardModif}). Indeed, the white-noise hypothesis has been introduced for the sake of simplicity but it is really far from clear if a turbulent velocity has something to do with a white noise. In \cite{Mallick04}, the effect of an exponential correlation of the noise (Ornstein-Uhlenbeck process) has been studied. It has been showed that the system behaviour was qualitatively the same: a stabilisation by the noise for weak noise intensities and a noise induced bifurcation for stronger noise. However, we know also that a turbulent velocity has not a pure exponential correlation function.

\begin{center}
\begin{figure}[h]
\includegraphics[scale=0.7,clip]{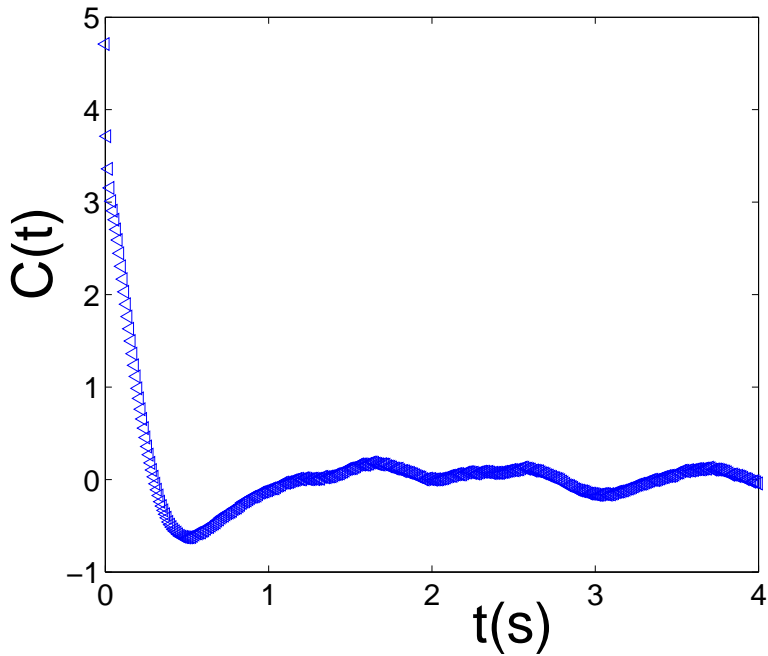}
\caption{\label{Correlation} Correlation function of the turbulent velocity from a von K\'arm\'an experiment.}
\end{figure}
\end{center}

\begin{center}
\begin{figure}[h]
\includegraphics[scale=0.7,clip]{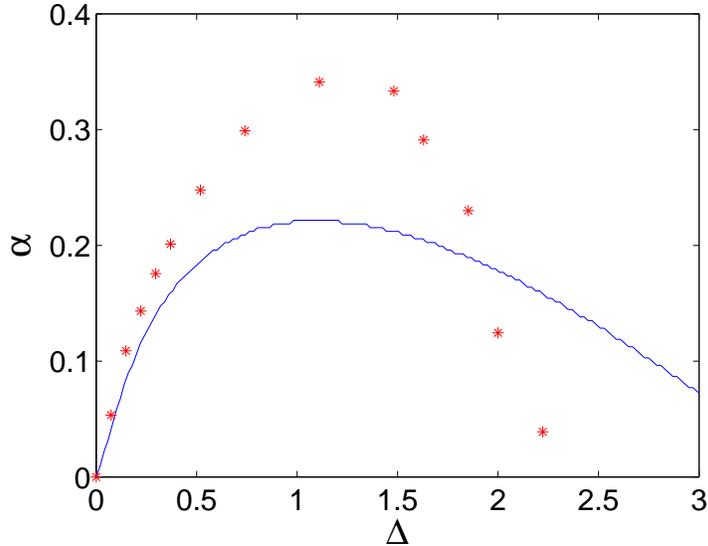}
\caption{\label{StabExp} Stability curve for the $\delta$ correlated model (straight line) and for the \lq\lq experimental noise\rq\rq (stars).}
\end{figure}
\end{center}

In order to get an insight into the influence of a \lq\lq real noise\rq\rq, we will use data from a fluid dynamics experiment. We will here use a signal of velocity from a von K\'arm\'an experiment. Details about the experimental setup can be found in \cite{Ravelet04} but for our purpose, it is enough to know that this device gives rise to a fully developed turbulence. On figure \ref{Correlation}, we show the correlation function that can be extracted from the fluctuating part of the velocity signal:
\EQ
c(t) = \frac{1}{T} \int_0^T v'(u) v'(u+\tau) du \qquad T \rightarrow \infty \; .
\EN
From the correlation function, we can define the parameters that will serve us to compare the correlated case, to the calculation we have made in the previous section. We will use the following definitions for the intensity of the fluctuation and the correlation time of the signal:
\EQ
\Delta = \int_0^\infty C(t) dt \qquad \text{and} \qquad \tau = \frac{\Delta}{C(0)} \; .
\EN
From the turbulent signal, we found a correlation time of $\tau \sim 0.14\; s$. By multiplying the turbulent fluctuating velocity $v'(t)$ by a suitable quantity, we can obtain a signal with any value for the parameter $\Delta$, while keeping the correlation time constant. Then, we use this signal in equation (\ref{Oscillateur}) as a noisy term ($\xi(t)=x'(t)$). Using the experimental noise, we observe as previously a succession of two bifurcations, first towards an intermittent state and then to a fluctuating one. However, the second transition is much more smooth than in the $\delta$-correlated case and is thus very difficult to identify. We will not discuss more about this transition. On the contrary, the first transition, from the absorbing state to the intermittent, is still easy to identify. We thus perform some numerical simulation, keeping $\Delta$ at a constant value and increasing the control parameter $\alpha$. The bifurcation line, separating an absorbing state (under the curve) and an intermittent state (above) is compared to the $\delta$-correlated case on figure \ref{StabExp}.
We see that the qualitative behaviour (stabilisation for weak noise and destabilisation for higher intensities) is not altered by the correlation of the noise. Moreover, we see that this scenario is reinforced: both the stabilisation and the destabilisation of the instability is much more marked than in the previous case.

\section{Summary and conclusions}
We studied a modified Bullard dynamo model showing that it can be mapped onto the problem of the stability of a non-linear oscillator in presence of a $\delta$-correlated noise. We studied the probability density of the position of the oscillator induced by the noise and showed that a non trivial scenario of bifurcation emerges because of the noise. First, looking at the mean value of the position, we found a transition from an absorbing state where the oscillator is trapped at his rest position for long time, to an intermittent state, where it stays close to the absorbing position for long time but sometimes makes excursion far from this position. However, the most probable position for the oscillator is still at the rest position. Thus, using the most probable value as an order parameter, we showed that a second bifurcation takes place for higher intensities of the control parameter. The evolution of the two bifurcation threshold has been studied relatively to the intensity of the noise. A recent result by \cite{Mallick04} has been recovered: for weak intensities, the instability is delayed whereas for higher noise, there is a reentrant transition induced by the noise. Furthermore, we showed that it was not the case for the threshold of the second transition which increases monotonously with the strength of the noise. Then, we checked what could be the influence of the correlation of the noise on this bifurcation scenario and showed that the qualitative picture was not changed but the threshold displacement was amplified by these correlations.

The transition to an intermittent state, known as the \lq\lq on-off intermittency \rq\rq in nonlinear physics has regained some attention in the past year because, despite its generality in chaotic system and stochastic system driven by multiplicative noise, there is very few experimental evidences of such a behaviour. In a recent paper, Aumaitre {\it et al.} \cite{Aumaitre05} have shown that the existence of intermittency was mainly monitored by the low-frequency components of the noise and because of filtering, these components may not be present in experimental devices. Regarding the dynamo effect, it is still a question if the first bifurcation will be towards an intermittent state ? The preceding study show that the intermittent bifurcation observed by \cite{Leprovost05b} is not an artefact's due to the white noise approximation. However, this behaviour could still be induced by other approximations such as the zero diffusivity limit. Furthermore, in a realistic dynamo experiment, there is other effects that could prevent us to observe a bifurcation towards an intermittent state, e.g. the presence of the external Earth magnetic field or the back-reaction of the magnetic field on the flow.

To study the effect of this back-reaction, we started the numerical study of a \lq\lq dynamical\rq\rq model, i.e. where the control parameter is not prescribed statistically but evolves with a chaotic equation of motion. In the kinematic approximation (where the effect of the magnetic field on the velocity field is neglected), the evolution of the magnetic field will be governed by the two first equation of (\ref{BullardModif}) with a chaotic variable $Z$. We aim to compare the results from the numerical study to the model presented here and investigate, firstly if the back-reaction of the magnetic field change anything in the bifurcation scenario and secondly what is the effect of a real correlated noise compared to the white one. This work is under progress.

\bibliographystyle{apsrev}
\bibliography{BiblioBullard2}

\end{document}